\documentclass[twocolumn,preprintnumbers,amsmath,amssymb,aps,pre]{revtex4}

\usepackage{latexsym}
\usepackage{dcolumn}
\usepackage{supertabular,multirow}
\usepackage{epsfig}
\usepackage{bm}
\usepackage{subfigure}
\usepackage{amsmath}
\usepackage{graphicx}
\usepackage{color}

\definecolor{brown}{rgb}{0.59, 0.29, 0.0}
 \definecolor{orange}{RGB}{255,127,0}
\definecolor{brightube}{rgb}{0.82, 0.62, 0.91}

\newcommand{\Ca}{\mbox{\it Ca}}

\newcommand{\Rr}{\mbox{\it R}}
\newcommand{\Sr}{\mbox{\it S}}

\newcommand{\zhat}{{\bf \hat z}}

\newcommand{\out}{{\mathrm{ex}}}
\newcommand{\ins}{{\mathrm{in}}}

\newcommand{\visrat}{{\lambda}}

\newcommand{\eps}{{\varepsilon}}

\newcommand{\bE}{{\bf{E}}}

\newcommand{\sigm}{{\sigma}}

\begin{document}

\title{Fluid rings and droplet arrays via rim streaming}

\author{Quentin Brosseau and Petia M. Vlahovska}
\affiliation{School of Engineering, Brown University, RI 02912, USA}

\date{\today}

\begin{abstract}

Tip-streaming generates micron- and submicron- sized droplets  when a thin thread pulled from the pointy end of a drop disintegrates. Here, we report streaming from  the equator of a  drop placed in a uniform electric field. The instability generates concentric fluid rings encircling the drop, which break up to form an array of  microdroplets in the equatorial plane. We show that the streaming results from an interfacial instability at the stagnation line of the electrohydrodynamic flow, which creates a sharp rim. The  flow draws from the rim a thin sheet which destabilizes and sheds fluid cylinders.  This  streaming phenomenon provides a new route for generating monodisperse microemulsions.

\end{abstract}

\maketitle

A highly conducting drop in a uniform electric field elongates into a prolate ellipsoid  whose  poles in strong fields  deform into conical tips (Taylor cones)  emitting jets of charged tiny droplets \cite{Basaran:1989, Collins:2013, Collins2008, delaMora, Khusid:2015}. This so called electrohydrodynamic (EHD) streaming or cone-jetting occurs in many natural phenomena (e.g., drops in thunderclouds) and technological applications (printing, electrospraying, electrospinning) \cite{delaMora, Basaran2013}. 

The streaming is related to a generic interfacial instability due to a convergent flow \cite{Prosperetti:2015},  see Figure 1.a.
The interface is compressed and a local perturbation at the stagnation point (e.g., drop tips)  gets drawn by the flow. If the viscous stresses overcome the interfacial tension,  the perturbation grows into a fluid filament. This is the tip-streaming phenomenon commonly observed 
 in the microfluidic co-flow geometry  \cite{Suryo:2006, Gordillo:2012, Anna:AR2016}. 
If instead of a point, the flow is converging to a stagnation line, then a thin sheet can be entrained \cite{Eggers:2001}. By analogy with the cone-jet geometry resulting from the destabilization of a stagnation point, it is expected that the instability of a stagnation line would give rise to an edge-sheet structure. In this Letter, we report for the first time streaming resulting from a stagnation line instability. 

Experimentally, we  exploit the electrohydrodynamic  flow about a neutral drop placed in a uniform electric field \cite{Taylor:1966, Melcher-Taylor:1969}. By varying the fluid conductivities, we are able create flow converging
 either at the drop poles  (Figure 1.b) to generate cone-jet, or at the equator ( Figure 1.c) to generate an edge-sheet. The latter case is the focus of this  work.

\begin{figure}[h]
 \centerline{\includegraphics[width=\linewidth]{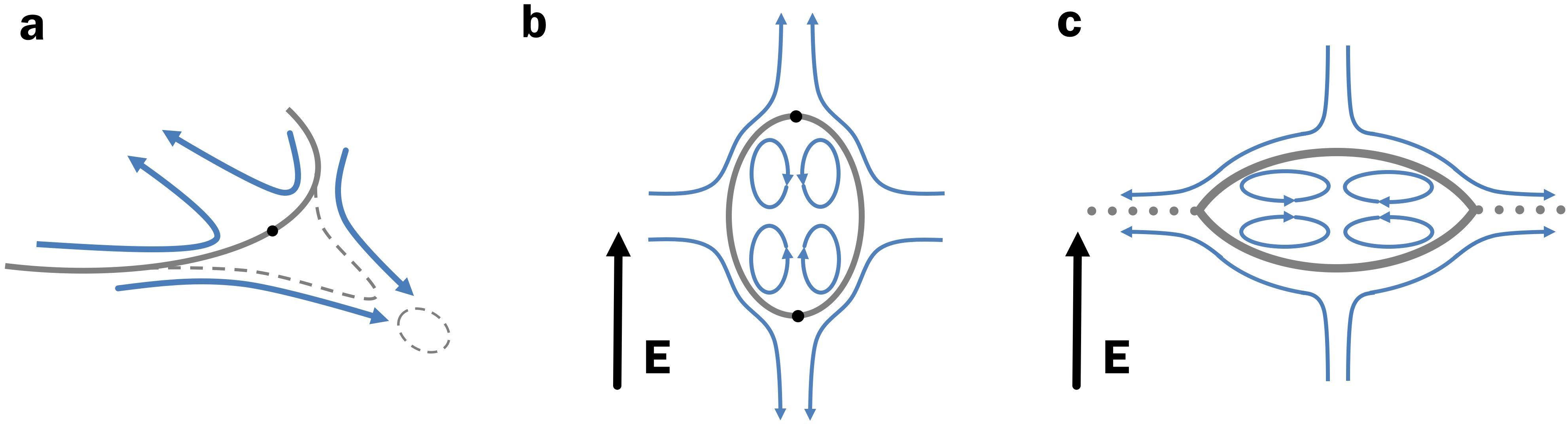}}
      \caption{\footnotesize (Color online) (a) A protrusion at the stagnation point of a convergent flow can grow into a filament. (b) For EHD flow about a highly conducting drop ($\Rr/\Sr>1$) the poles are stagnation points where cone-jets form. (c) For EHD flow about a low conducting drop ($\Rr/\Sr<1$) the equator is a stagnation line where edge-sheet is expected to form.}
      \label{fig1}
      \end{figure}

The electrohydrodynamic flow is driven by electric shear stresses due to induced surface charges\cite{Taylor:1966, Melcher-Taylor:1969}. 
For a  drop in a uniform electric field the resulting flow is axisymmetrically aligned with the applied field. For a spherical drop with radius $a$ placed in DC electric field $\bE=E\zhat$, the surface velocity is \cite{Taylor:1966}
\begin{equation}
\label{svel}
{\bf{u}}_T
=\frac{a \eps_\out E^2}{\mu_\out}\frac{9(\Sr-\Rr)}{10(1+\visrat) (\Rr+2)^2} \sin(2\theta)\hat{\bm{\theta}}, 
\end{equation}
where $\visrat=\mu_\ins/\mu_\out$ is the viscosity ratio between the drop  and suspending fluids and $\theta$ is the angle with the applied field direction.

The direction of the surface flow depends on the  difference of conductivity, $\sigma$,  and permittivity, $\eps$, of the drop and suspending fluids $\Rr=\sigm_\ins/\sigm_\out$ and $\Sr=\eps_\ins/\eps_\out$.  For highly conducting drops,  $\Rr/\Sr>1$, the surface flow is from the equator to the poles. Accordingly,  the poles become stagnation points where streaming occurs at sufficiently strong fields,  see Figure 1.b. Since the tips are also the location of maximum induced charge ($Q\sim \sin \theta$), the emitted drops carry away some of it and become charged.

If the drop is less conducting than the suspending medium, $\Rr/\Sr<1$, the surface flow is from the pole to the equator. Here the equator is a stagnation line. {\it{Could streaming occur in such geometry? What structures are formed? }} 
Surprisingly, drop stability under such conditions has been studied only to a very limited extent. The experimental \cite{Torza} and  numerical studies \cite{Lac-Homsy, Ghazian} showed that the drop  dimples at the poles and becomes a torus  \cite{Zabarankin:2013}.  A streaming instability related to the equatorial stagnation line as expected by the flow convergence  \cite{Prosperetti:2015} has not yet been observed.

\begin{figure}[h]
	\centering
	\includegraphics[width=\linewidth]{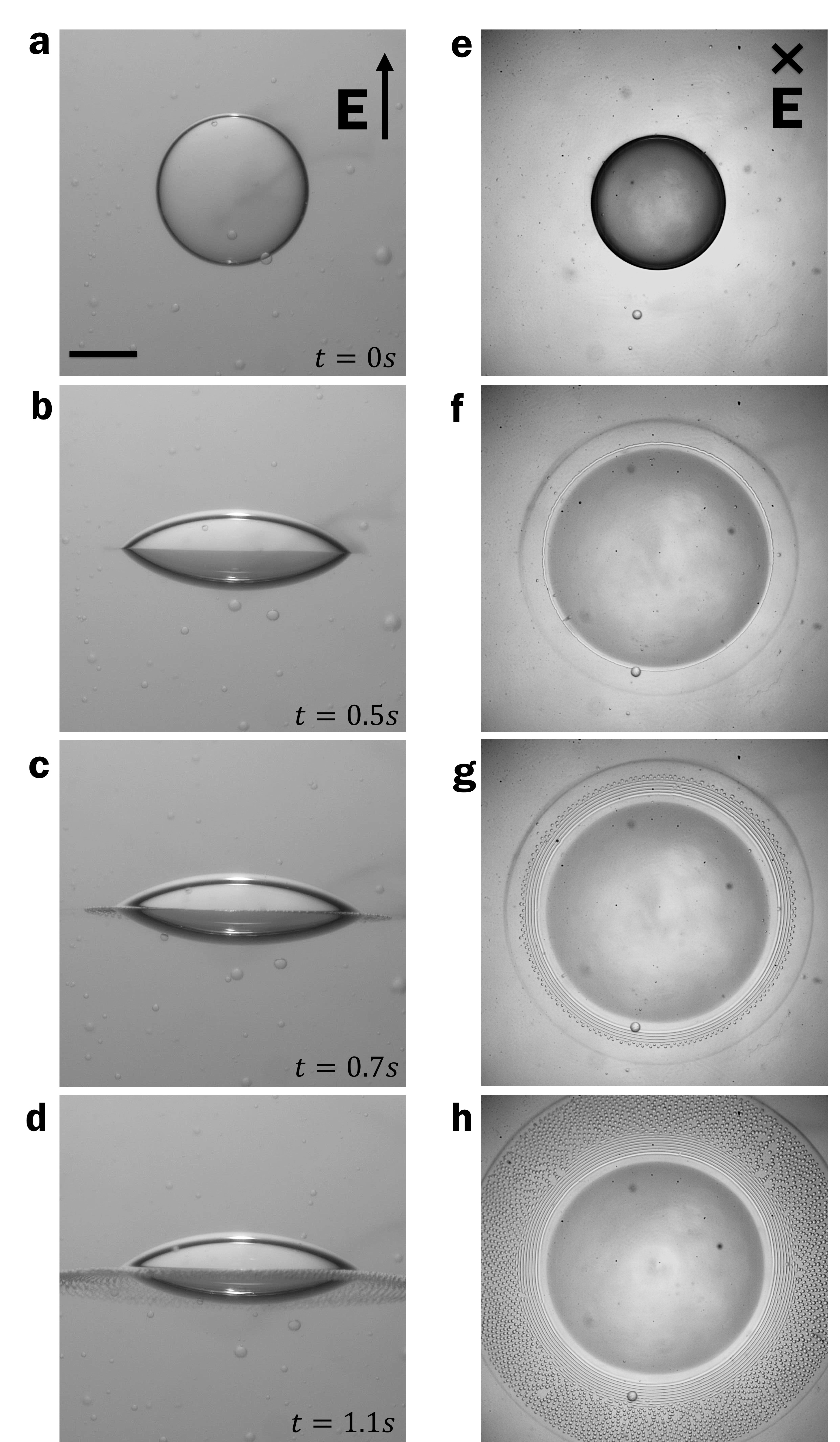} 	
	\caption{\footnotesize  Development of the rim instability  observed  from direction perpendicular to the applied electric field (a-d), and along the applied  electric field (e-h); the field direction is the axis of symmetry. Spherical drops  (a,e)  deform as the electric field ($E=4kV/cm$) is turned on at $t=0$. In equatorial-streaming the mother drop flattens to aspect ratio of about 0.5 and forms a sharp, cusp-like rim (b)-(f). The emission of rings occurs radially in the equatorial plane of the drop (g)-(h).Viscosity ratio is $\lambda=0.07$. Scale bar is 500$\mu m$.}
\label{fig2}
\end{figure}

{\em{Experiment:}} The fluid system and experimental set up are similar to \cite{Salipante-Vlahovska:2010}.
Silicone oil (SO) and castor oil (CO) are used as drop and suspending fluids, respectively. Both fluids have low conductivity (in the order of $10^{-12}\, S/m$) and high viscosity (100 to 1000 times that of  water).  CO viscosity is $ \mu_\out=0.69 \, Pa.s$ and SO viscosity is varied to adjust the viscosity ratio $\lambda=\mu_\ins/\mu_\out$ in the range between 0.001 to 10.  For this system, the permittivity ratio is  $\Sr=\eps_\ins/\eps_\out=0.6$  and the conductivity ratio  $\Rr=\sigm_\ins/\sigm_\out$ is varied between $5\times10^{-6}$  and $3\times 10^{-2}$ by doping the CO with organic electrolyte
TBAB (Tetrabuthylamonium Bromide) or  AOT (Dioctyl sulfoccinate sodium salt). The surface tension $\gamma$  in all cases is measured to be 4.5 mN/m confirming that the TBAB and AOT are not surface-active in the SO/CO system. A uniform DC electric field is generated in a rectangular chamber built around two parallel ITO coated glass electrodes, both 75x50 mm and set 25mm apart.
In the experiment, a millimeter-sized drop is pipetted manually in the middle of the chamber, far away from any boundary.
The field is then turned on and the voltage is increased stepwise in increments of approximately  0.1kV/cm; at each step the system is allowed to equilibrate in order to avoid spurious transients. Drop deformation/destabilization is recorded by CCD cameras placed either perpendicular to electric field, or parallel to it (through one electrode).
Figure \ref{fig2} illustrates the phenomenon of equatorial  streaming as recorded in each two directions. 

  \begin{figure}[h]
	\centering
	\includegraphics[width=\linewidth]{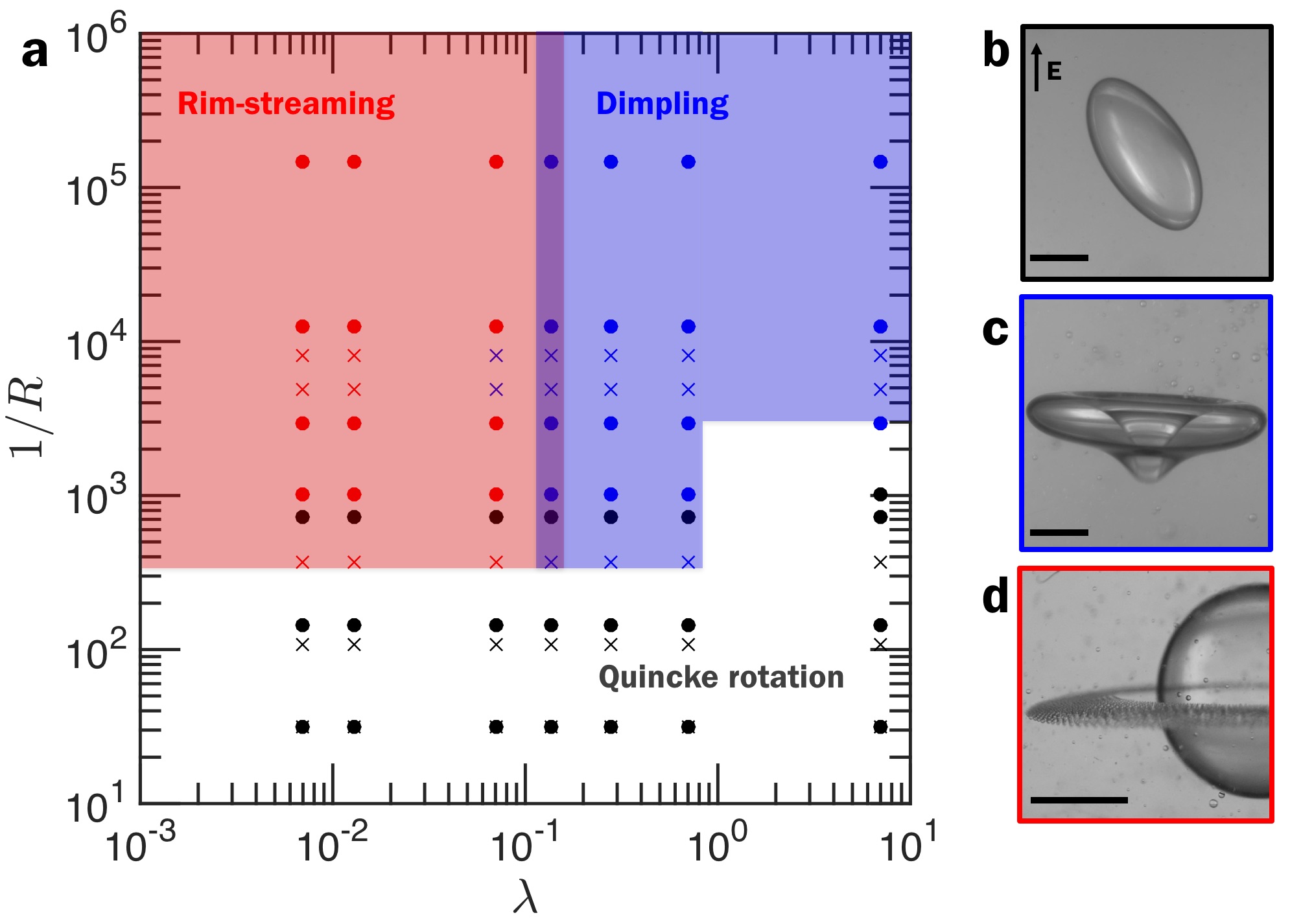} 	
	\caption{\footnotesize (Color online)  {{Phase diagram of the drop dynamics in a strong uniformDC electric field}}:
	Quincke electrorotation (black) for $E\ge 2.7 kV/cm$, dimpling (blue)  for $\lambda>0.1$ and $E\ge 2.3kV/cm$ , equatorial rim-streaming (red) for  $\lambda<0.1$ and $E\ge 3-5kV/cm$. The conductivity of the suspending oil is modified by addition of electrolytes TBAB  (dots) or AOT  (crosses). Scale bar 500 $\mu m$. Images (b)-(d) illustrate the drop modes.}
\label{fig3}
\end{figure}

{\em{Results:}}  
The classic leaky-dielectric  theory by G. I. Taylor \cite{Taylor:1966, Melcher-Taylor:1969} predicts that in weak electric fields, $\Ca=\eps_\out E^2 a/\gamma\ll 1$,  and if $\Rr/\Sr<1$  a drop adopts an oblate spheroidal shape, the flow and shape being axisymmetrically aligned with the applied field. As the field strength increases, the drop undergoes various types of instabilities depending on fluids viscosities and conductivities. Figure \ref{fig3}.a maps the modes of droplet destabilization as a function of fluids properties. 
There are three distinct modes:

(A) {\it {Electrorotation:}} In this regime, the drop tilts relative to the applied field direction, see Figure \ref{fig3}.b. This symmetry-breaking is due to the Quincke electrorotation \cite{Salipante-Vlahovska:2010, Vlahovska:PRFreview, Yariv:2016}, which gives rise to a rotational flow about the drop. The Quincke effect stabilizes the drop against break-up and even decreases interface deformation \cite{He:2013}. Further increase in the field strength leads to various time-dependent behaviors (e.g., tumbling) \cite{Salipante-Vlahovska:2013}. This mode is observed for any viscosity ratio and low conductivity of the suspending fluid  (which corresponds to $\Rr \lesssim 1$). The threshold for electrorotation, $E_Q$ is
estimated  from the value for a rigid sphere \cite{Jones:1984}
 \begin{equation}
\label{EQ}
 E_Q^2=\frac{2\sigm_\out \mu_\out \left(\Rr+2\right)^2}{3\eps_\out ^2 (\Sr-\Rr)}\,.
\end{equation}
For the pure fluid system $E_Q=2.7kV/cm$. Adding electrolytes to the suspending fluid increases its conductivity, $\sigm_\out$, by several orders of magnitude and shifts the  transition to Quincke to higher field strengths thereby effectively suppressing the electrorotation. In the absence of electrorotation, the following two  modes of drop fragmentation emerge.

 (B) {\it {Dimpling:}} In this mode, the drop deforms into a biconcave disc with rounded rim  and pinches in its center to form a torus, see Figure \ref{fig3}.c; the torus subsequently breaks into few drops  \cite{Torza, Ghazian, Pairam:2009}. The drop burst is  abrupt, uncontrollable, and the resulting daughter-droplet size and number is rather irreproducible. This mode creates few drops with size comparable to the mother drop. It is observed for moderate to high  viscosity ratios $\visrat>0.1$. 
The dimpling break-up occurs at field strengths about 2.3kV/cm  corresponding to capillary number  $\Ca\sim O(1)$, when the distorting electric stresses can no longer be contained by the interfacial tension. 
Intriguingly, similar deformation and break-up behavior is reported in numerical simulations of a drop in compressional flow \cite{Zabarankin:2013}.

\begin{figure}[h]
	\centering
	\includegraphics[width=\linewidth]{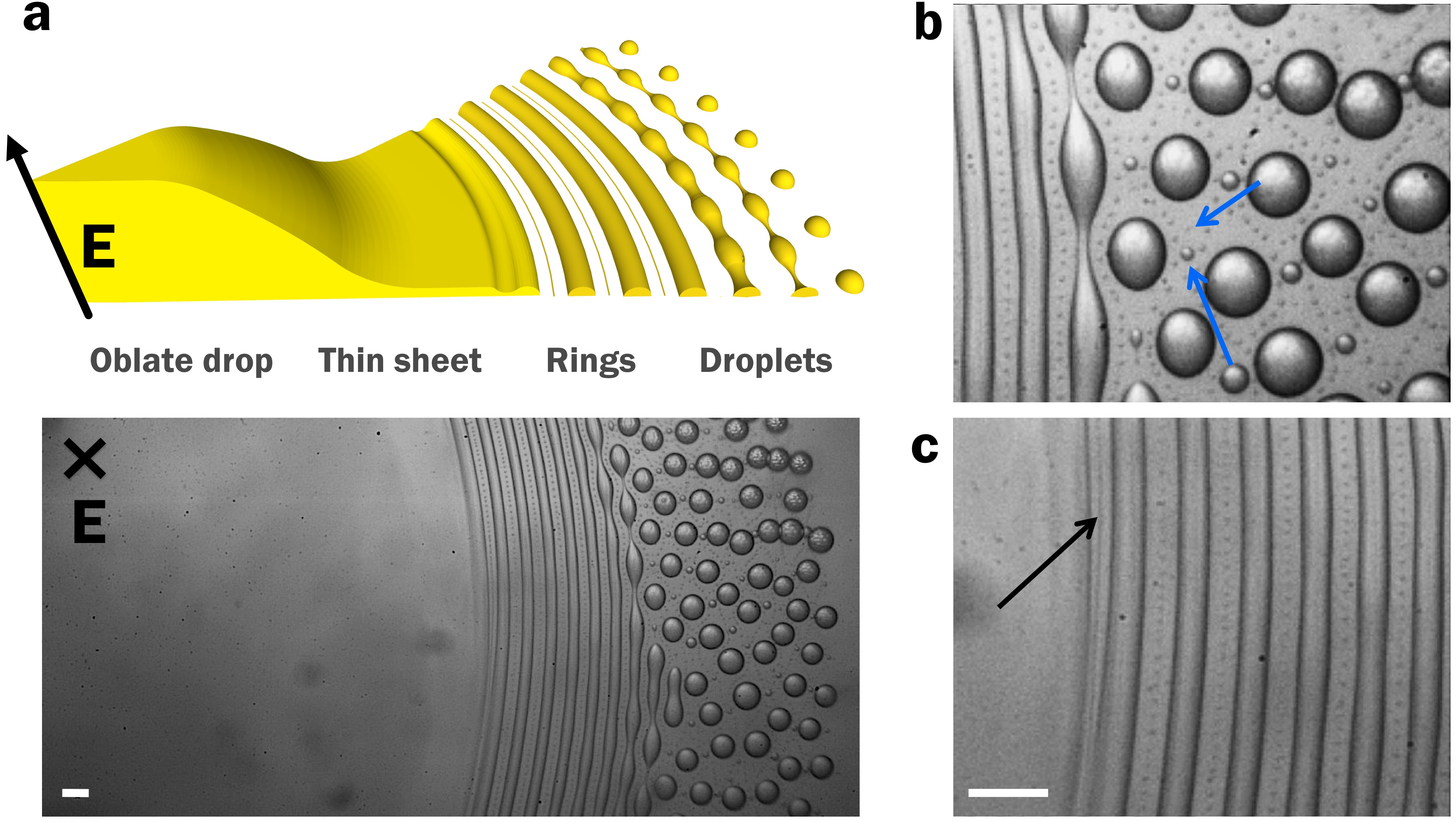} 	
	\caption{\footnotesize (Color online) Rim-streaming (a) 3D rendering of the phenomenon as deduced from the  experimental image ($\lambda=0.07$). The thin sheet is visible as a slightly brighter region just prior the edge where the ring is forming. A closer inspection reveals (b) two generations of satellite drops created after the ring break-up  (blue arrows) and  (c)  a satellite ring that itself breaks up in droplets (black arrow). Scale bars 100$\mu m$.}
\label{fig4}
\end{figure}

\begin{figure}[th]
	\centering
	\includegraphics[width=3.5in]{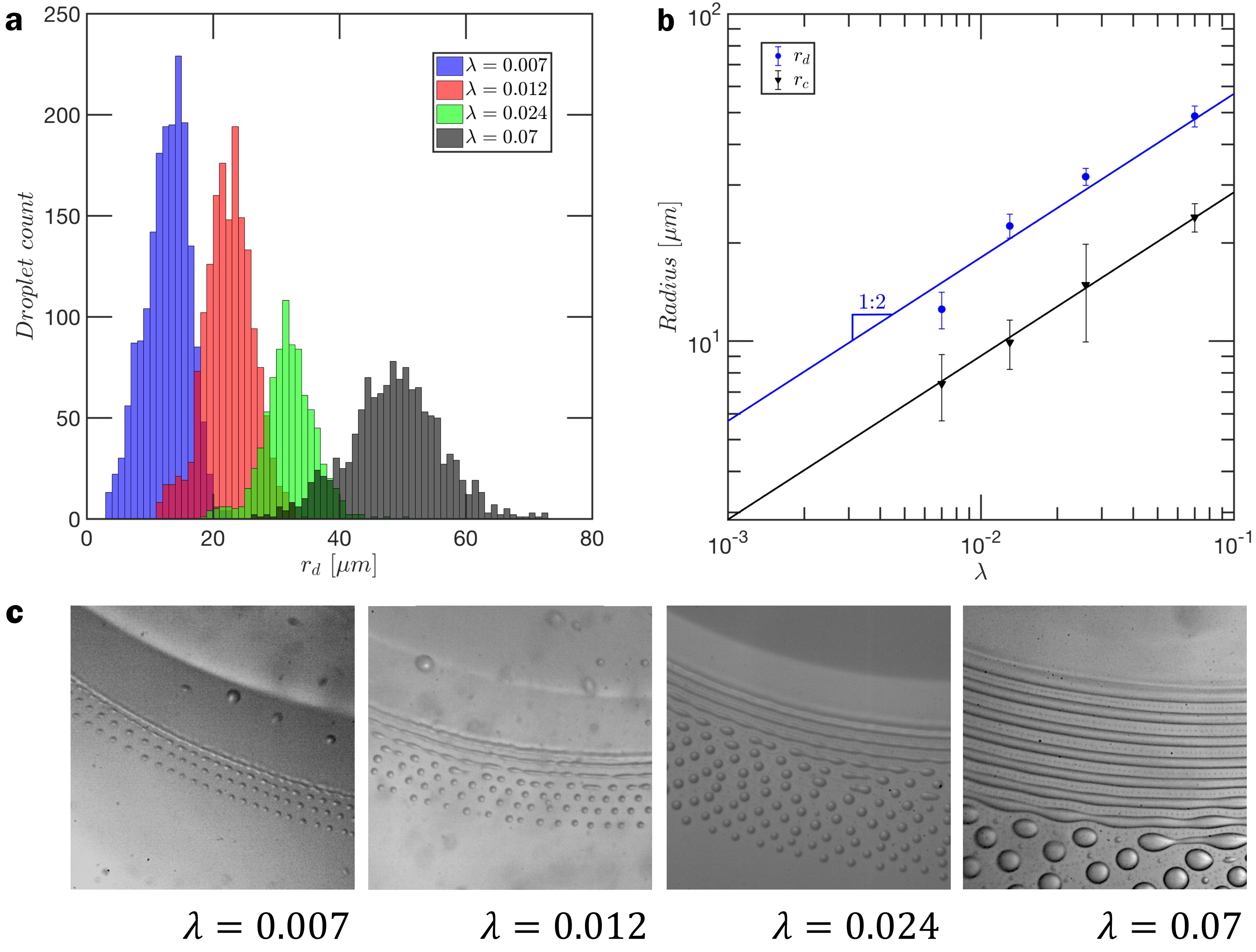} 	
	\caption{\footnotesize (Color online) (a) Droplet size distribution at different viscosity ratio $\visrat$. Standard deviation varies between $15-30\%$ of the central value. (b) Ring $r_c$ and droplet $r_d$ radii follow $ \sim \lambda^{1/2}$ dependence on viscosity ratio.
 (c) The cylinder size and lifetime increase with the viscosity ratio, changing the morphology of the streaming.}
\label{fig5}
\end{figure}

(C)  {\it {Rim streaming:}} In this mode the drop flattens and forms a sharp rim with thin film attached to it (edge-sheet), see Figure \ref{fig2}. The sheet sheds  concentric fluid rings with diameter 2-3 orders of magnitude smaller than the mother drop; the rings subsequently break-up via capillary instability into microdroplets, see Figure \ref{fig4}. The ring shedding occurs in a steady manner  so droplet production can proceed  for  tens of seconds. Unlike dimpling the streaming is a controllable process that is easily triggered and interrupted, e.g., see Figure \ref{fig3}.c which illustrates a mother drop surrounded by daughter droplets after the field is turned off. 
This streaming mode is able to produce thousands of microdroplets with relatively uniform size distribution, see Figure \ref{fig5}.a. 
 
The equatorial rim-streaming phenomenon occurs via a multistep process involving a downsizing cascade from one macro-drop, to a thin edge-sheet, to concentric fluid rings, to thousands of micro-droplets, see Figure \ref{fig4}.
This mode is observed only for low viscosity drops,   $\visrat<0.1$, and at field strengths corresponding to capillary number  $\Ca\sim O(10)$ significantly higher than the one required for ``dimpling''  break-up. If  we define  $\Ca_L=\eps_\out E^2 L_c/\gamma\sim O(1)$ as the condition for break-up then in the case of dimpling the characteristic length scale $L_c$ is the drop radius  $L_c\sim a$, since this breakup involves the whole drop. In the case of streaming, $\Ca\sim O(10)$ implies  a smaller characteristic length scale $L_c\sim a/10$ most likely related to the scale of the equatorial edge-sheet, where the breakup cascade occurs.  
 
In a first step, the millimeter-sized drop flattens to aspect ratios of about 0.5 and forms a sharp rim at the equator. A thin-sheet extends for few hundreds of microns around the drop equator.  The process can be qualitatively explained by the instability  recently analyzed by Tseng and Prosperetti \cite{Prosperetti:2015}.  A perturbation on the                                       interface gets entrained by the converging flow to  form a filament or sheet. The surface tension opposes the interface deformation and needs to be weakened for the instability to grow. In the classical tip-streaming, surfactant convected to the stagnation point lowers the surface tension. In EHD cone-jetting, the induced surface charge at the tips  reduces the surface tension.  The situation in our system is more complex because  surfactant is absent and the stagnation line is in fact the location of zero induced charge, e.g., for a sphere the induced charge varies as $\cos\theta$.  
However, all variables undergo sharp variations near the equator (the stagnation line) \cite{supplem}. The induced  charge changes polarity and as the drop becomes spheroidal the transition  gets sharper  compared to a sphere. What is directly relevant to our problem is that the electric pressure increases significantly compared to a sphere and becomes more localized near the equator. This localized electric pressure pulls on the interface and overcomes  the surface tension. Moreover, the shear electric stresses which drive the convergent flow strengthen as the aspect ratio increases. Both effects - stronger electric pressure and shear stress aid the growth of the instability.  
This mechanism likely explains the upper limit of viscosity ratio 0.1 for this process to happen. This value is  comparable to the viscosity threshold for tip-streaming in hydrodynamic focusing \cite{Anna:AR2016}.

In a second step, the sheet pinches off at the leading edge and sheds rings with a typical radius $r_c$ about $20\mu m$, see Figure \ref{fig4}.  A thin sheet (unlike cylinder) is stable in the context of capillary instability, so the destabilization must be due to the flow (similarly to hydrodynamic focusing flows) or the electric field. 

In a final step, the rings break into droplets. The breakup wavelength agrees well with the prediction by the capillary instability theory for a cylinder. The concentric rings break up via an out-of-phase correlation between neighboring rings: due to hydrodynamic interactions an alternation of necking and expanding occurs along the orthogonal direction \cite{Zhang:SM2015}, see Figure \ref{fig4}.c.  
 
The capillary instability is also characterized by the formation of satellite droplets during breakup. The ratio of droplet diameter from generation to mother is a function the viscosity ratio \cite{Tjahjadi1992}. For $\lambda=0.07$, the daughter-mother size ratios of the two generations visible in our experiments, indicated by the blue arrows on Figure \ref{fig4}.c, is 
about 0.2 and 0.1, in good agreement with the numerical predictions.
One generation of a satellite cylinder is also created as a ring detaches from the edge-sheet, see Figure \ref{fig4}.d. (black arrow). For $\lambda=0.07$, the size-ratio is $\sim0.2$. The secondary cylinder undergoes faster capillary break up, as the thread lifetime is a function of the diameter.

Figure \ref{fig5} shows that the microdroplets generated via rim streaming are quite uniform in size. Their size increases with the viscosity ratio as $\visrat^{1/2}$.
This power-law dependency seems to originate from the slenderness of the sheet from which the cylinders are formed.  According to slender body theory for a thin film of length $L$ and thickness $d\sim 2r_c$,  the balance of viscous shear stresses imposed by the external flow, $\mu_\out U/L$,  and lubrication pressure in the thin film, $\visrat \mu_\out U L/d^2$ yields $d/L\sim \visrat^{1/2}$ \cite{Stone:1994}. Since the rings break-up via capillary instability, droplet size is set by the cylinder radius and follows the same dependence of viscosity ratio.  

The cylinder radius directly affects its lifetime. As the viscosity ratio decreases the rings get thinner and break-up faster. Accordingly,  the number of concentric rings decreases, see Figure \ref{fig5}.c. For example,  a ring is barely visible at $\lambda=0.007$ and the droplet seem to originate directly from the edge-sheet. This also suggests that at even lower viscosity ratios, rings may not observed.

{\em{Concluding remarks:}}  In this Letter we report that upon application of a uniform DC electric field,  a drop flattens, forms a sharp rim with a thin film attached to it (edge-sheet) shedding charge-free fluid rings  encircling the drop. The concentric fluid rings subsequently undergo capillary instability and break up into droplets. The droplets form an initially  hexagonal pattern in the equatorial plane of the mother drop.  The streaming occurs only for low viscosity drops, with viscosity ratio smaller than 0.1 and field strengths corresponding to $\Ca\sim O(10)$.

While the mechanism of the streaming is yet to be quantified, the phenomenon is qualitatively explained by the interfacial instability of the stagnation line of a convergent flow \cite{Prosperetti:2015}. The flow is driven by electric shear stresses on the drop interface and converges at the equator. A perturbation of the compressed interface grows and a fluid sheet is drawn from the equator, which is the stagnation line. The growth of the interface deformation into an edge-sheet structure is aided by he normal electric stresses which overcome the surface tension. 

Rim-streaming allow the production of large droplet quantity in a relatively short time. The final droplet size can be tuned by changing the viscosity ratio. This study suggests ways of microdroplets production in bulk environment,  ``electroemulsification'', with potential application in industrial processes. 

We hope our experimental observations will inspire further work into this phenomenon. Numerical simulations are needed to explain the  effects of viscosity ratio and field strength in the selection of the ``streaming'' versus the ``dimpling'' mode of drop destabilization. The generic nature of the instability suggests that  rim streaming can be obtained  in absence of electric field, for example, a surfactant-covered drop in axisymmetric compressional flow.  

{\em Acknowledgements:} This research was supported by the NSF - CBET 1437545  and CMMI-1538703 awards.

\bibliographystyle{unsrt}

\end{document}